\documentclass[runningheads]{llncs}
\usepackage[T1]{fontenc}

\usepackage{subfigure}
\usepackage{graphicx}
\usepackage{float}
\usepackage{color}

\definecolor{myblue}{RGB}{0, 123, 236}
\definecolor{mypink}{RGB}{235, 220, 247}
\definecolor{mylightblue}{RGB}{211, 237, 253}

\begin{document}
\title{ContextVis: Envision Contextual Learning and Interaction with Generative Models}
\titlerunning{ContextVis}
%
\author{Bo Shui\orcidID{0009-0001-0981-9608} \and 
Chufan Shi\orcidID{0009-0005-7889-5187}\and \\Yujiu Yang\orcidID{0000-0002-6427-1024}\and Xiaomei Nie\thanks{Corresponding Author.}\orcidID{0000-0002-8901-8981}}
\authorrunning{Bo Shui et al.}

\institute{Shenzhen International Graduate School, Tsinghua University, Shenzhen, China
\email{\{shuib22, scf22\}@mails.tsinghua.edu.cn}\\
\email{\{yang.yujiu, nie.xiaomei\}@sz.tsinghua.edu.cn}}
\maketitle

\begin{abstract}
ContextVis introduces a workflow by integrating generative models to create contextual learning materials. It aims to boost knowledge acquisition through the creation of resources with contextual cues. A case study on vocabulary learning demonstrates the effectiveness of generative models in developing educational resources that enrich language understanding and aid memory retention. The system combines an easy-to-use Dashboard for educators with an interactive Playground for learners, establishing a unified platform for content creation and interaction. Future work may expand to include a wider range of generative models, media formats, and customization features for educators.
\end{abstract}

\section{Introduction}
Recently, generative models have made significant strides in the field of artificial intelligence, demonstrating remarkable capabilities in creating realistic and coherent content across various media formats. These models, such as language models and image generation models, have been widely adopted in the creation of creative resources, offering unique opportunities to enrich materials with contextual information. The deployment of generative models in educational resource development holds considerable promise for augmenting the learning process. It provides learners with an immersive framework that aids in knowledge comprehension and retention. 

In this regard, we present the ContextVis system, a workflow that harnesses the capabilities of generative models to craft tailored learning materials and experiences. Through a case study on vocabulary learning, we demonstrate the utility of generative models in generating educational content that supports language development and enhances learning outcomes. The system represents a promising approach to elevate contextual and exploratory learning using generative models and interactive visualizations.

\section{Related Work}
\subsection{Leveled and Themed Language Learning}
Leveled reading, a pedagogical approach characterized by evaluating children's reading skills using specialized instruments and subsequently supplying reading materials commensurate with their ability, has garnered international acceptance and has continued to evolve in recent years \cite{scholasticLeveledReading}. Optimal selection of reading content that presents an appropriate challenge is crucial to fostering advancement in reading proficiency \cite{wang2023preliminary}.

Empirical evidence suggests that the use of leveled reading materials in English for primary school students learning English as a second language can substantially improve their linguistic capabilities, comprehension skills, and cultural awareness \cite{wu2017empiricalreading}. Renowned publications such as \textit{Reading A-Z} \cite{ReadingAZOnline}, \textit{Geronimo Stilton} \cite{philosophygeo}, and \textit{Oxford Reading Tree} \cite{OxfordReadingTree} adeptly integrate repetitive sentence structures and captivating illustrations that resonate with the developmental stages of a child’s cognition \cite{li2017reasearchinto,nurhayati2023efficiency}.

Theme and context play a vital role in language learning. The idea of contextualized learning emphasizes that knowledge is best acquired within the context in which it is used and applied \cite{brown1989situated}. Learners apply their knowledge in various themes and contexts, deconstructing complex concepts, identifying relationships between elements, and adeptly engaging in problem analysis and resolution \cite{kloos2021educational,shui2023community}. For effective learning, the language input should be comprehensible, engaging, relevant, and abundant, focusing more on meaning rather than on grammatical rules \cite{krashen1989we}.

The recent \textit{English Curriculum Standards for Compulsory Education} issued by the Ministry of Education of China in 2022 advocate for English language teaching through leveled and contextual exploration, offering level guidelines and theme examples \cite{ministry2022englishcurriculum}. Our work utilizes the standards as a foundation for leveled vocabulary and theme references to develop a system that supports both teachers and students in the language instruction.

\subsection{Generative Models for Education}
Recent advancements in generative artificial intelligence models have been propelled by the growth of available data and model scaling. These models now exhibit capability to generate high-quality content across multi-modalities and demonstrate improved understanding of user instructions, enabling content creation that aligns with human intentions \cite{cao2023comprehensive}. Notably, empirical evidence has suggested that large generative models are now able to synthesize emergent content that is not only contextually pertinent but also coherent, contingent upon specific prompts \cite{wei2022emergent}.

The integration of generative models in the creation of educational resources offers a unique opportunity to enrich learning materials with contextual cues, thereby enhancing the learning experience. They have so far been used in various educational applications, such as storytelling co-creation with specific knowledge \cite{zhang2024mathemyths}, generating reading comprehension quizzes \cite{dijkstra2022reading}, authoring data-driven articles \cite{sultanum2023datatales}, generating coherent comics \cite{jin2023generating}, and creating stories \cite{alabdulkarim2021automatic,shi2024lifi}. Moreover, the collaboration with human teachers in the creation of educational resources has been shown to be effective in enhancing the quality and alleviating teachers' workload \cite{ji2023systematic}. 

These studies demonstrate the potential of generative models in creating educational resources that are tailored to specific learning objectives and contexts, as well as integrating generation with human teachers to enhance the quality of the contents. Our work focuses on maintaining the consistency of the learning context in generated story scripts and images for the learning vocabulary.

\subsection{Interaction Design for Human-AI Collaboration}
Human-AI collaboration is the study of how humans and AI agents work together to accomplish a shared goal \cite{sturm2021coordinating}. In order to understand and improve this emergent modality of interaction, HCI researchers have proposed a variety of principles, frameworks, and guidelines in recent years \cite{amershi2019guidelines,cila2022designing,cimolino2022two,lubart2005can}. Notably, Amershi et al. have proposed 18 guidelines for Human-AI Interaction, categorized within the sequential phases of initial stage, active interaction, error response, and post-interaction reflection \cite{amershi2019guidelines}. Cimolino et al. proposed a framework as an analysis tool with four dimensions of the role, supervision, influence and mediation of AI \cite{cimolino2022two}. 

The domain of AI-assisted design, particularly in graphical and interaction design, has witnessed a significant increase in interest and application since 2016 \cite{shi2023understanding}. 

Shi et al. proposed that AI can assist designers in Discovering, Visualizing, Creating and Testing, and designers can reciprocally augmenting AI by Training and Regulating \cite{shi2023understanding}. In the teaching process, educators harness AI assistance in lesson planning, thereby liberating themselves from repetitive pedagogical tasks, which in turn fosters innovation in educational content and methodologies \cite{zhang2023reasearchpath}. The guidelines for Human-AI interaction laid out in the literature are highly pertinent to our work, as we focus on leveraging AI in the creation, visualization, and validation phases of educational content generation.

\section{The ContextVis System}
\begin{figure}[h]
\includegraphics[width=\textwidth]{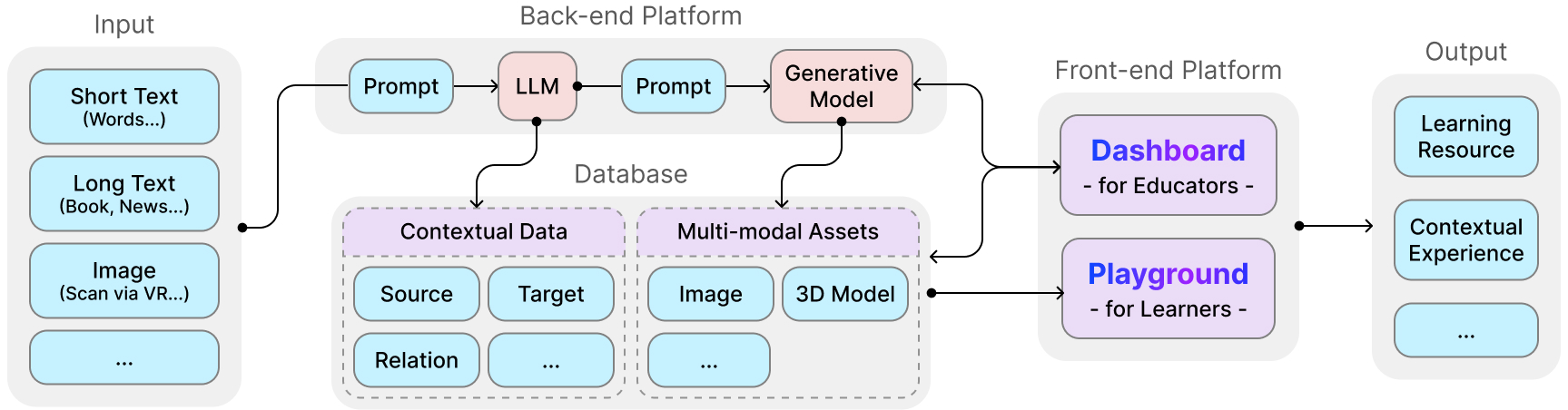}
\caption{The general workflow of ContextVis system.} \label{fig:workflow}
\end{figure}

We present the ContextVis system, which leverages generative AI to develop contextual learning materials and experiences. As illustrated in Fig. \ref{fig:workflow}, the workflow comprises three essential components: the back-end platform for processing user input and generating contextual data using Large Language Models (LLMs), as well as generating multi-modal assets through generative models; the database, where the metadata and generated resources are stored; and the front-end platform, featuring a Dashboard for educators and a Playground for learners. It is worth noting that our vision for this system encompasses the ability to receive inputs and generate outputs in multiple media formats, as the models employed can be substituted to accommodate diverse tasks.

\section{Case Study: Contextual Vocabulary Learning}
Acquiring English vocabulary can be challenging due to its inherent complexity characterized by contextual nuances, semantic diversity, and an extensive lexicon. Understanding and memorizing grammar rules alongside vocabulary demands a coherent and persistent learning strategy, often requiring prolonged exposure to a variety of linguistic stimuli. This extended engagement with both language inputs and outputs is crucial for reinforcing linguistic knowledge. 

\begin{figure}[h]
\includegraphics[width=\textwidth]{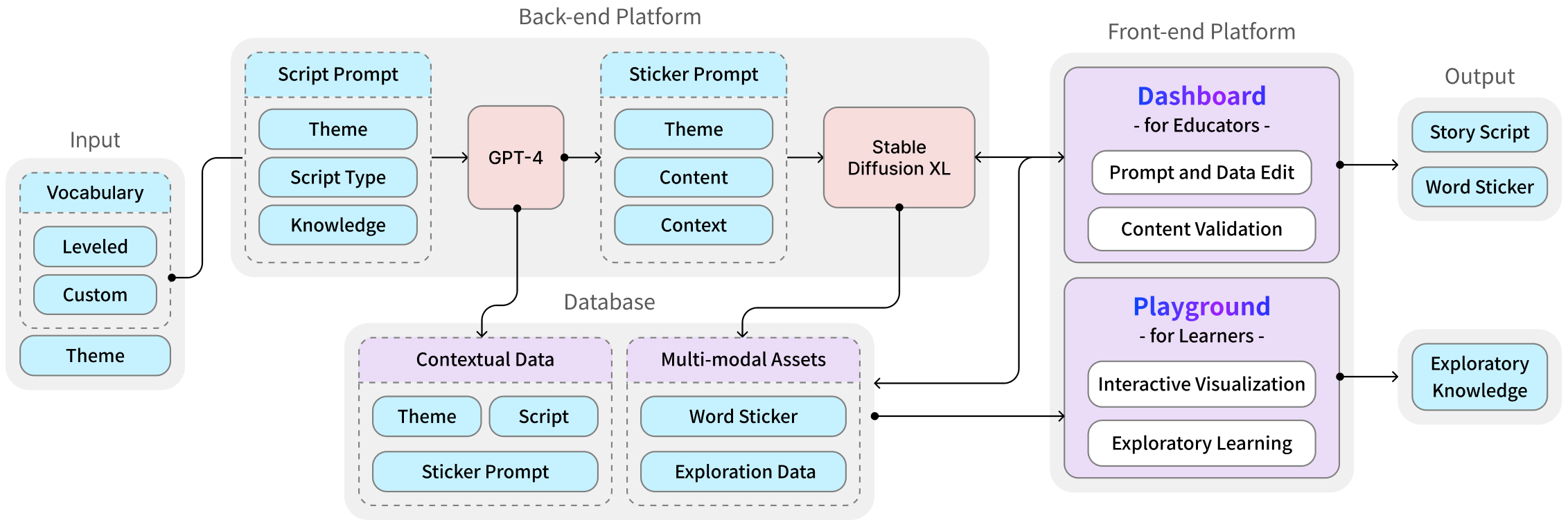}
\caption{A case workflow of ContextVis system.} \label{fig:case_workflow}
\end{figure}

The process of vocabulary acquisition can be effectively integrated into the proposed workflow, where the integration of contextual factors plays a crucial role in fostering language development. To this end, we conducted a case study for vocabulary learning which exemplifies the significance of the ContextVis system, shown in Fig. \ref{fig:case_workflow}. In the workflow, with selected vocabulary in a unit and an optional theme input in the Dashboard, the generative models in the back-end automatically generates a contextually coherent a story script and stickers for each word as output. The generated data and assets are stored in the database on the server for learners to access and explore in the Playground.

\subsection{Dashboard for Educators}
\begin{figure}[htb]
    \centering
    \includegraphics[width=\textwidth]{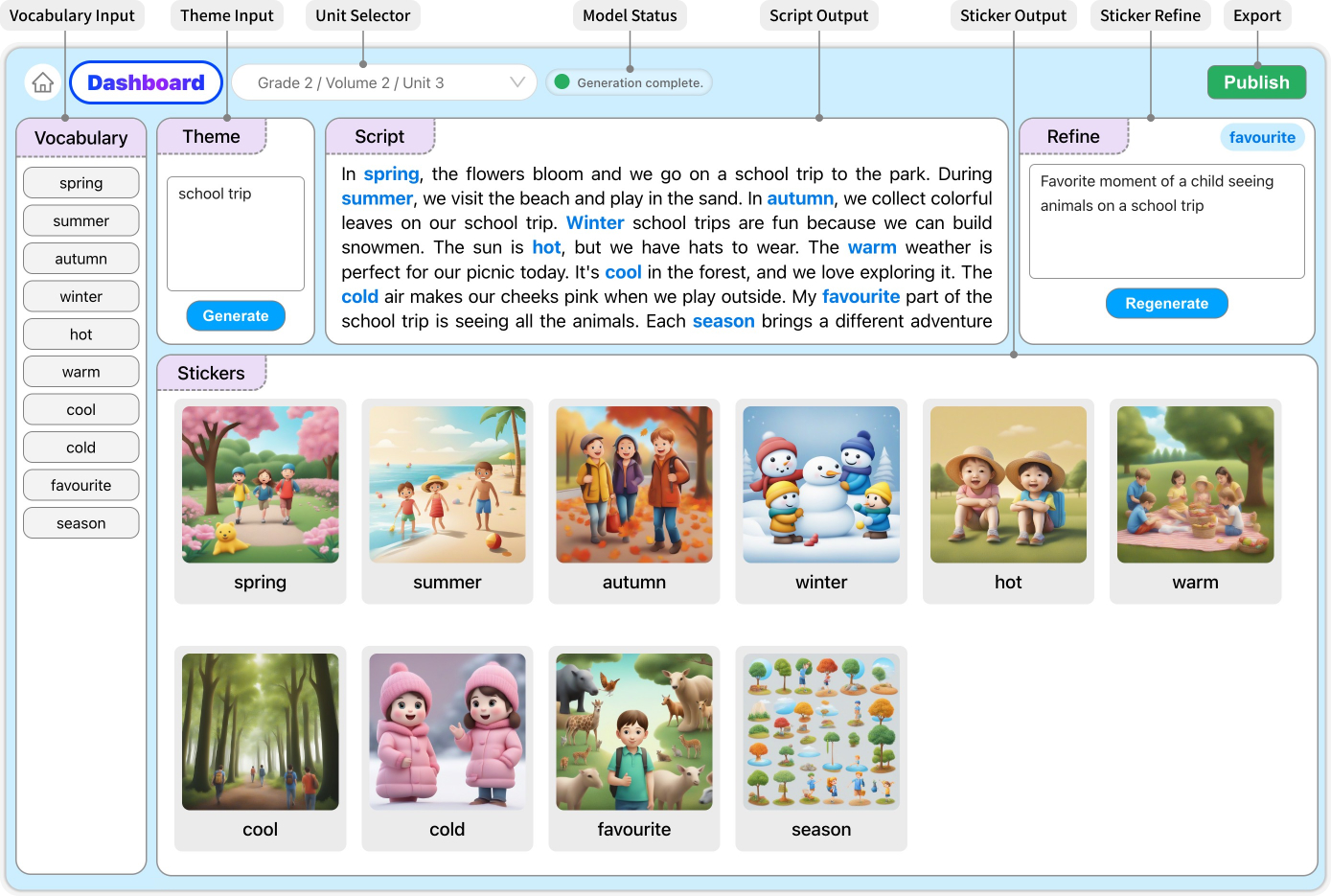}
    \caption{An overview of the Dashboard.}
    \label{fig:dashboard}
\end{figure}

To facilitate seamless integration of vocabulary with generative contents, we developed a Dashboard for educators, shown in Fig. \ref{fig:dashboard}, with multiple panels allowing educators to input and select the vocabulary, then generate, preview and refine the generated educational resources. By assigning a theme as contextual cue, the generated assets of scripts and stickers are stored in the server and can be exported as teaching materials for classroom use. The Dashboard serves as a pivotal interface within the ContextVis system, offering user-friendly tools and features that empower educators to effectively create tailored educational content.

\vspace{-2em}
\subsubsection{Story Script} Stories serve as powerful media for contextualizing vocabulary, integrating words organically within cohesive narratives. We extracted vocabulary from English textbooks used in primary schools in China to generate story scripts using the state-of-the-art large language model, GPT-4 \cite{openai2023gpt}. Educators select vocabulary and input in the \colorbox{mypink}{\textsf{Theme}} panel to send request for generation on the server. A model status indicator is shown in the header next to the unit selector to inform the process of generation on the server. When the script is generated, it is displayed in the \colorbox{mypink}{\textsf{Script}} panel with the words highlighted to show their contextual usage. The prompt template and an example script is shown below, in which the words are from a Grade 2 textbook for primary schools:
\\
\colorbox{mylightblue}{\textsc{Prompt:}}
\texttt{Generate sentences consisting of the following words with the theme of \textcolor{myblue}{\textit{\textbf{<school trip>}}}, make sure the sentences are easy and s\\hort for primary school students learning English as a second lang\\uage, one sentence for one word and all sentences are contextually \\related to form an article. And generate a prompt for each word ba\\sed on the sentence for image generation in stable diffusion. Here \\are the words: \textcolor{myblue}{\textbf{\textit{<spring, ..., cool>}}}.}
\\
\\
\colorbox{mylightblue}{\textsc{Output:}}\\
\texttt{
        Word: 
        \textcolor{myblue}{\textit{\textbf{"spring"}}}\\
        Sentence: 
        \textcolor{myblue}{\textit{\textbf{"In spring, the flowers bloom and we go on a school trip \\to the park."}}}\\
        Sticker Prompt: 
        \textcolor{myblue}{\textit{\textbf{"Children on a school trip in a park full of bloom\\ing flowers representing spring."}}}\\
        ...\\
        Word: 
        \textcolor{myblue}{\textit{\textbf{"cool"}}}\\
        Sentence: 
        \textcolor{myblue}{\textit{\textbf{"It's cool in the forest, and we love exploring it."}}}\\
        Sticker Prompt: 
        \textcolor{myblue}{\textit{\textbf{"A group of students exploring a cool, shaded fore\\st."}}}
}

\subsubsection{Word Sticker}  
To further enhance the expressiveness of the generated story script, we incorporated word stickers into the narrative. The Stable Diffusion XL (SDXL) model \cite{podell2023sdxl} with two Low-Rank Adaptation (LoRA) modules \cite{hu2021lora} is employed to accomplish the task, where the LoRA modules fine-tune the SDXL base model to create emoji-like images based on the sticker prompts \cite{noauthor_sdxl_2023} and accelerate the generation \cite{luo2023lcm}. Using the sticker prompts generated with the script, example of which shown in the \colorbox{mylightblue}{\textsc{Output}} section above, the system generates stickers of coherent theme for each word. These stickers are subsequently transmitted to the Dashboard, populating the \colorbox{mypink}{\textsf{Stickers}} panel. 


\subsubsection{Data Refine and Theme Variants}
To encourage divergent thinking and foster customization, educators can further generate variant scripts and utilize alternative prompts, expanding the possibilities for tailoring the output content to specific learning objectives. Upon selecting an item in the \colorbox{mypink}{\textsf{Stickers}} panel, the current prompt associated with the chosen sticker is displayed in the \colorbox{mypink}{\textsf{Refine}} panel. Educators can edit the prompt and regenerate stickers to better fit the intended context, while simultaneously mitigating the risk of generating potentially inappropriate content.

\begin{figure}[h]
\includegraphics[width=\textwidth]{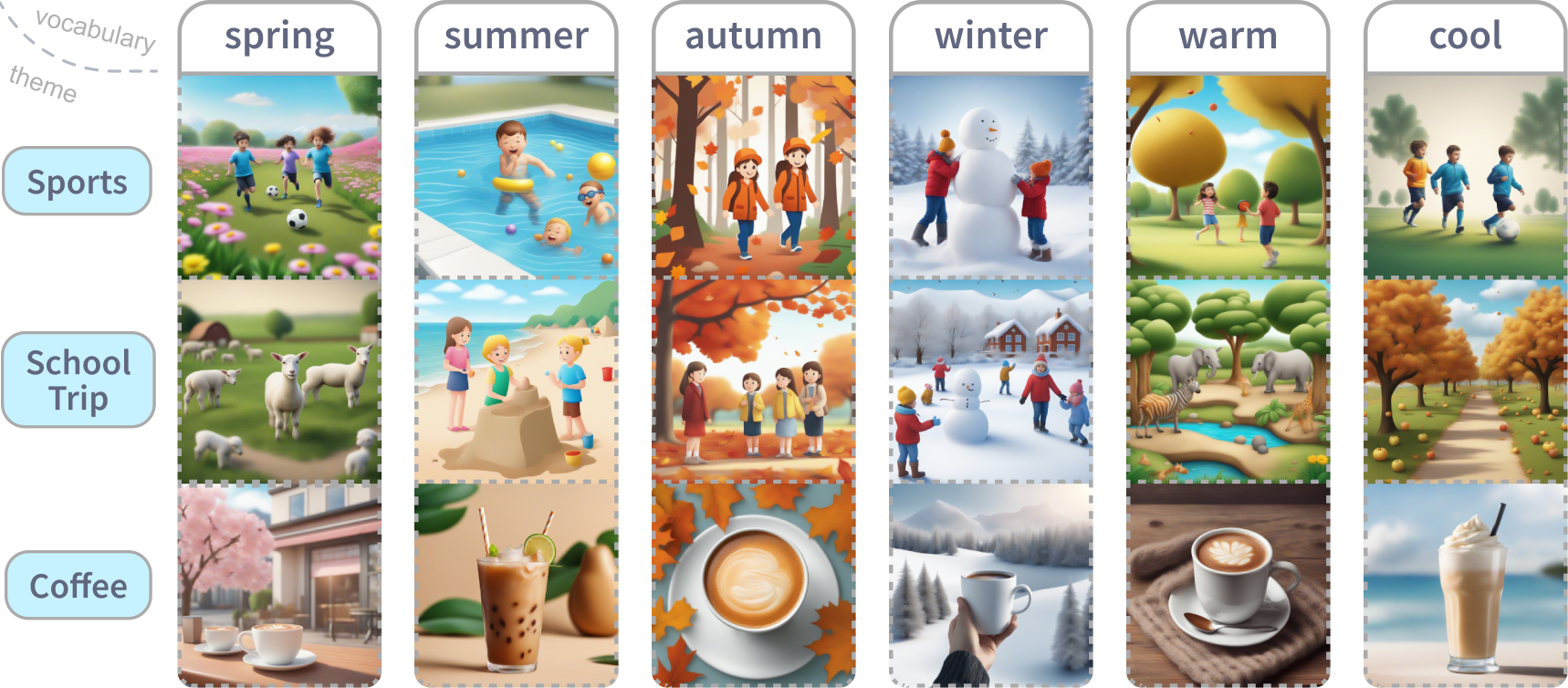}
\caption{Stickers generated for the same vocabulary in different contexts.} \label{fig:sticker_example}
\end{figure}

Moreover, by leveraging varied themes and contexts, educators can create more engaging learning contents that not only reinforce vocabulary acquisition but also promote linguistic adaptability, as shown in Fig. \ref{fig:sticker_example}. By offering a diverse array of learning materials, educators can stimulate more dynamic forms of vocabulary engagement and deepen learners' connections with the language.

\subsection{Playground for Learners}
\begin{figure}[h]
    \centering
    \includegraphics[width=\textwidth]{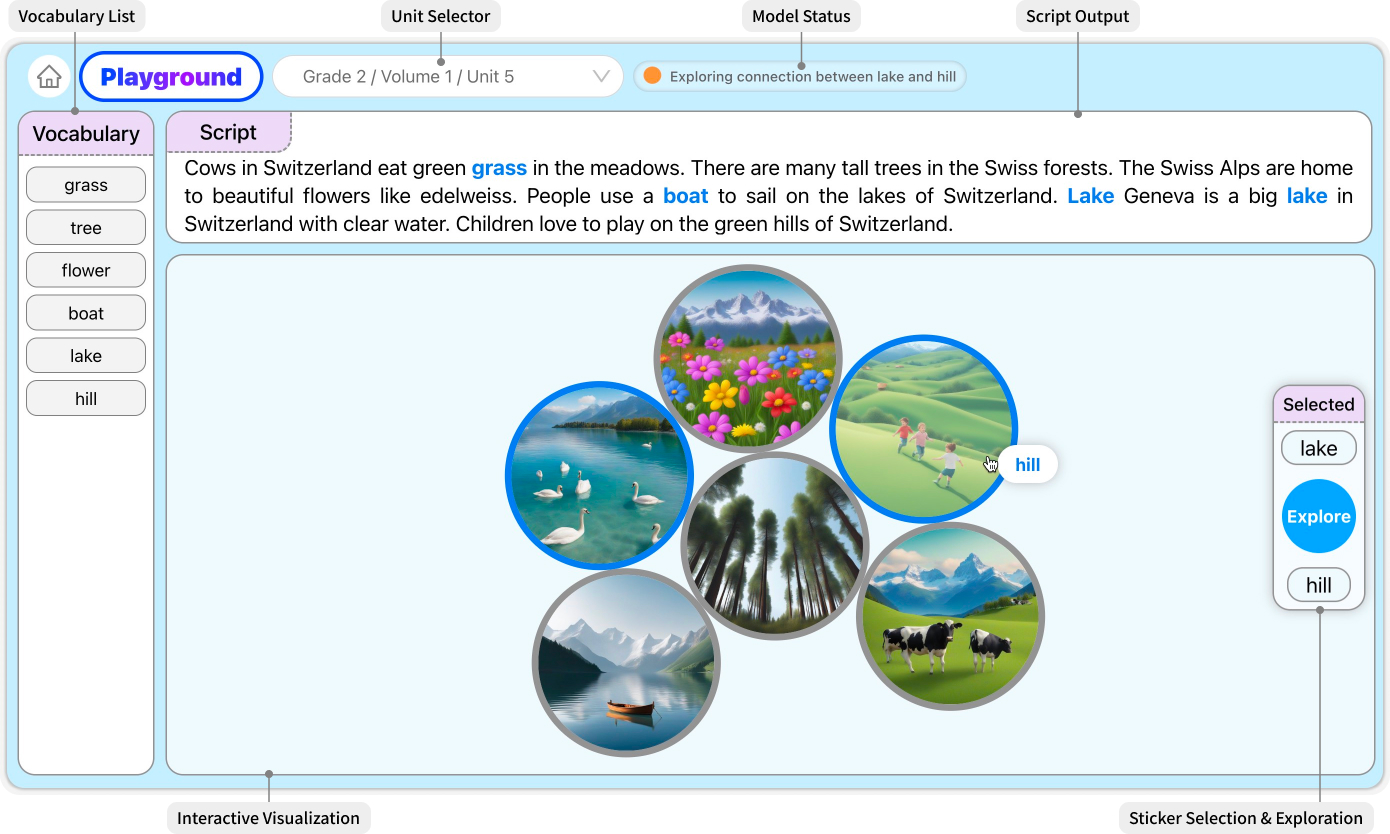}
    \caption{Explorer view of the Playground.}
    \label{fig:playground_overview}
\end{figure}

In addition to the generated learning contents, the Playground serves as a platform for learners to engage in interactive exploration. This interactive space offers learners the opportunity to actively review with the concepts and vocabulary introduced in the generated content and to explore more by themselves. It consists of four panels: Vocabulary, Script, Selected and the interactive visualization in the center, as shown in Fig. \ref{fig:playground_overview}. The Playground is seamlessly synchronized with the Dashboard as the two platforms are interconnected by the data and assets they share in the database on the server. This integration ensures that the learning materials sophisticatedly prepared by educators are available for learners to review and explore extra contents aligning with the theme of the original materials.

\vspace{-0.5em}
\subsubsection{Interactive Visualization}
Learners select the unit in the header, and the Playground send queries to the database to retrieve the corresponding script and stickers. The \colorbox{mypink}{\textsf{Script}} and \colorbox{mypink}{\textsf{Visualization}} panels are updated with incoming data, so the learner can interact with the stickers to explore the connections between them. The visualization utilizes the D3.js \cite{D3ObservableJavaScript} library to create an interactive network, where the stickers attract each other to mimic the context, aiding learners in comprehending the contextual relationship between words and their corresponding usage within the script.
\vspace{-0.5em}
\subsubsection{Exploratory Learning}
\begin{figure}[h]
    \centering
    \includegraphics[width=\textwidth]{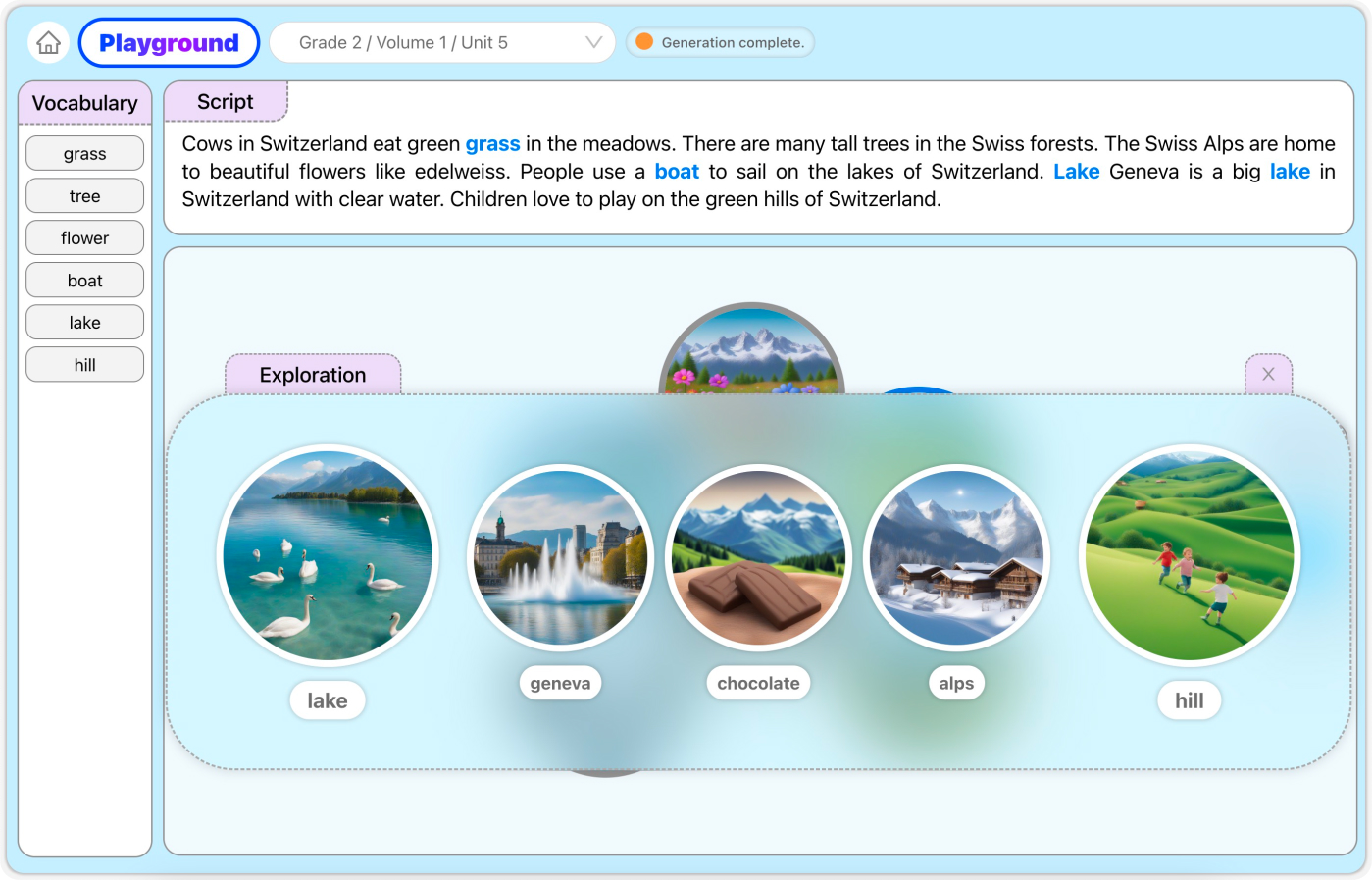}
    \caption{Example of a relation explored in the Playground.}
    \label{fig:playground_connection}
\end{figure}

The visualization additionally facilitates a deeper investigation of the interconnections between vocabularies. When the stickers are selected, they are highlighted and appear in the \colorbox{mypink}{\textsf{Selected}} panel, where a maximum of two stickers can be selected at the same time, prompting the user to explore further into the relationship between them. Upon clicking the Explore button, additional vocabulary and stickers will be generated under the same context, to establish a connection between the selected items in a pop-up \colorbox{mypink}{\textsf{Exploration}} panel, as shown in Fig. \ref{fig:playground_connection}. These newly generated elements either act as supplementary materials for the original resources or diverge from the original learning materials, thereby fostering divergent thinking and exploratory learning.

The prompt template and an example output is shown below:
\\
\\
\colorbox{mylightblue}{\textsc{Prompt:}}
\texttt{
Find the relationship between the following words based on the semantic relevance with the theme of \textcolor{myblue}{\textit{\textbf{<Switzerland>}}}. Add rel\\ated words to make the semantic relevance more consistent. Add a p\\rompt for each word based on the theme for image generation in sta\\ble diffusion. Ensure the first and last two words are the input w\\ords, for example: "Monitor, Mouse" - "Monitor-Computer-Computer A\\ccessories-Mouse". Here are the two input words: \textcolor{myblue}{\textbf{\textit{<lake, hill>}}}.
}
\\
\\
\colorbox{mylightblue}{\textsc{Output:}}\\
\texttt{
        Word: 
        \textcolor{myblue}{\textit{\textbf{"geneva"}}}\\
        Sticker Prompt: 
        \textcolor{myblue}{\textit{\textbf{"Cityscape of Geneva, Switzerland, with the iconic \\Jet d'eau fountain and lake Geneva in the foreground."}}}\\
        Word: 
        \textcolor{myblue}{\textit{\textbf{"chocolate"}}}\\
        Sticker Prompt: 
        \textcolor{myblue}{\textit{\textbf{"Swiss chocolate bars with the Swiss alps mountain \\in the background."}}}\\
        Word: 
        \textcolor{myblue}{\textit{\textbf{"alps"}}}\\
        Sticker Prompt: 
        \textcolor{myblue}{\textit{\textbf{"The majestic Swiss alps on a sunny day, with pict\\uresque ski resorts and chalets"}}}
}

\section{Discussion}
In this article, we presented the ContextVis system, a workflow that leverages generative models for creating contextual learning materials and experiences. Through a case study on contextual vocabulary learning, we have demonstrated the utility of generative models in resource creation and the effectiveness of the system in fostering language development. 

The user-friendly Dashboard for educators and an interactive Playground for learners together provide a comprehensive platform for both content creation and engagement. The integration of generative models, contextual cues, and interactive visualizations offers unique opportunities for learners to understand, explore, and retain knowledge. During the development stage, we found that providing competent information such as theme and context to generative models is essential to generate high-quality learning materials.

Future directions for the ContextVis system include expanding the range of generative models and media formats, as well as more in-depth customization options for educators. Overall, the ContextVis system represents a promising approach to enhance contextual and exploratory learning using generative models and interactive visualizations.

\subsubsection{Acknowledgements} This work was supported by a research grant from Shenzhen Key Laboratory of Next Generation Interactive Media Innovative Technology (Funding No: ZDSYS20210623092001004) and the Center for Social Governance and Innovation at Tsinghua University, a major research center for Shenzhen Humanities \& Social Sciences Key Research Bases.

%
\bibliographystyle{splncs04}
\bibliography{mybib}
\end{document}